\documentclass[journal]{IEEEtran}
\usepackage{amsmath}
\usepackage{graphicx}
\usepackage{subfigure}
\usepackage{booktabs}
\usepackage{color}
\usepackage{mathrsfs}
\usepackage{makecell}
\usepackage{gensymb}
\usepackage{multirow}
\usepackage{cite}
\usepackage{array}
\newcommand{\PreserveBackslash}[1]{\let\temp=\\#1\let\\=\temp}
\newcolumntype{C}[1]{>{\PreserveBackslash\centering}p{#1}}
\newcolumntype{R}[1]{>{\PreserveBackslash\raggedleft}p{#1}}
\newcolumntype{L}[1]{>{\PreserveBackslash\raggedright}p{#1}}

\usepackage[colorinlistoftodos]{todonotes}

\begin{document}

\title{\bf A Two-Stage Wavelet Decomposition Method for Instantaneous Power Quality Indices Estimation Considering Interharmonics and Transient Disturbances}

\author{Yiqing~Yu,~Wei~Zhao,~Shisong~Li,~\IEEEmembership{Senior Member,~IEEE,}
        and~Songling~Huang,~\IEEEmembership{Senior Member,~IEEE}
\thanks{This manuscript has been accepted for publication in {\it IEEE Transactions on Instrumentation and Measurement} and the copyright has been transferred to IEEE. Y. Yu, W. Zhao, and S. Huang are with the State Key Laboratory of Power System, Department of Electrical Engineering, Tsinghua University, Beijing 100084, China (e-mail: wyq16@mails.tsinghua.edu.cn; zhaowei@mail.tsinghua.edu.cn; huangsling@mail.tsinghua.edu.cn). S. Li is with the Department of Engineering, Durham University, Durham DH1 3LE, U.K. (e-mail: leeshisong@sina.com).}
\thanks{This work was supported by the National High-tech R$\&$D Program of China (863 Program, Grant No. 2015AA050404) and the Transforming Systems through Partnership (Grant No. TSPC1051) from the
Royal Academy of Engineering, UK.}}

\markboth{}{}

\maketitle

\begin{abstract}
As the complexity increases in modern power systems, power quality analysis considering interharmonics has become a challenging and important task. This paper proposes a novel decomposition and estimation method for instantaneous power quality indices (PQIs) monitoring in single-phase and three-phase systems with interharmonics and transient disturbances. To separate the interharmonic components, a set of new scaling filter and wavelet filter with narrow transition bands are designed for the undecimated wavelet packet transform (UWPT). Further, a two-stage decomposition method for multi-tone voltage and current signals is proposed. The Hilbert transform (HT) is applied to calculate the instantaneous amplitude and phase of each frequency component, which accordingly allows the monitoring of different PQI parameters. Numerical tests are conducted to check the performance of the proposed method. The test results show that compared to other conventional approaches, instantaneous PQIs estimated by the proposed method present significant advances for tracking transitory changes in power systems, and could be considered as a helpful tool for high-accuracy PQ detections.
\end{abstract}

\begin{IEEEkeywords}
Hilbert transform (HT), instantaneous power quality indices, interharmonic, wavelet filter, transient disturbance, undecimated wavelet packet transform (UWPT).
\end{IEEEkeywords}

\IEEEpeerreviewmaketitle

\section{Introduction}
\IEEEPARstart{I}{n modern} electricity market, monitoring and controlling the power quality (PQ) is an important concern for both electricity utilities and users \cite{1}. In the past few decades, the increasing number of nonlinear loads, random switching operations of large electrical loads, and the wide use of power electronic devices have become the major sources of PQ disturbances, e.g., impulsive transients, oscillatory transients, interruptions, sag, harmonic distortion, interharmonics, etc. Without a doubt, under such situations, strong nonstationary properties exhibit in the waveforms of voltage and current, which leads the PQ analysis a more difficult task than it was in the past \cite{2,3,4,5}.

Generally, the detection, localization, and classification of PQ disturbances, as well as the quantification of power distortions, are mainly based on the continuous monitoring of power quality indices (PQIs) \cite{3,6}. In this regard, IEC 61000-4-30 \cite{7} gives the parameters for PQ event detection, IEC 61000-4-7 \cite{8} provides indices for harmonics and interharmonics measurement, and IEEE 1459 \cite{9} defines a series of PQIs which can give comprehensive analysis of PQ in both single-phase and three-phase power systems. These PQIs are defined mainly based on the fast Fourier transform (FFT). However, FFT, with its extended version short-time Fourier transform (STFT), can only give accurate results under stationary conditions, where the magnitudes and frequencies of the spectral components of voltage and current signals are stable within a certain period. While analyzing the nonstationary signals, severe spectral leakages may appear in the spectra of FFT and STFT \cite{10,11}. STFT, compared to FFT, can access the time-related information by reducing the window length, which, however, decreases the frequency resolution unavoidably \cite{12}.

The wavelet transform (WT) is proved to be an effective tool for PQI estimations \cite{6}, \cite{13,14,15,16}. In \cite{15}, the power and RMS quantities were redefined in the time-frequency domain using the discrete wavelet transform (DWT) for the first time. Inspired by \cite{15}, single-phase and three-phase PQIs were reformulated using the wavelet packet transform (WPT) \cite{6,16}. Both DWT and WPT can provide a more accurate PQ analysis result than FFT or STFT when considering nonstationary signals. Nevertheless, both these redefinitions used the WT coefficients obtained in a time-window of 0.2\,s, which limits their application for fast PQ transient disturbance detection. 

Regarding the continuous monitoring of PQIs, instantaneous indices have been proposed in several previous papers. In \cite{17}, a frequency-shifting DWT decomposition algorithm, so-called FS-DWT, and the Hilbert transform (HT) were employed to analyze the instantaneous frequency characteristics of electrical power waveforms. { The FS-DWT method overcomes the spectra leakage problem in the discrete wavelet packet transform and hence has a higher detection accuracy for PQIs than the previous method.} In [18], instantaneous PQIs of single-phase systems were redefined based on time-frequency distribution (TFD) and cross time-frequency distribution (XTFD) of transient voltage and current signals. { The technique can preserve both time and frequency information in transient signals, introducing a new approach for instantaneous power component estimations according to IEEE 1459.} In [19], instantaneous amplitudes, frequencies, and phases of both even and odd harmonics of voltage and current signals were estimated simultaneously using WPT in accordance with frequency-shifting  (FS-WPT) and HT, and the instantaneous PQIs were calculated using these instantaneous values instead of Fourier series coefficients in the conventional PQIs definitions in IEEE 1459. These estimated instantaneous PQIs are able to capture the ``transient'' characteristics of disturbance signals. However, all the above researches do not consider the existence of interharmonics \cite{20,21}. Particularly, since the HT considers that the analyzed signal is single-tone, FS-DWT and FS-WPT methods may lead to large estimation errors in systems with interharmonics. PQIs that depend on the interharmonic contents cannot be calculated accurately.

This paper aims to seek solutions for estimating instantaneous PQIs in power systems with interharmonics and transient disturbances.  One of the major challenges of solving this issue is to separate each frequency component in voltage and current signals, especially the interharmonic components that are close to the fundamental component and integer harmonics. In this work, the undecimated wavelet packet transform (UWPT), which can be seen as an improved version of WPT with an additional feature of time invariance, is employed. A set of new scaling filter and wavelet filter of UWPT with narrow transition bands are designed. They are used to separate the interharmonics from the fundamental frequency component and integer harmonics. After the residual harmonic components of voltage and current signals are decomposed by conventional wavelets, the instantaneous PQIs are finally estimated using the HT. 

This paper is a significantly improved version of \cite{22} and achieves two advanced properties: 1) The effect of filter ripples is suppressed. The two-stage decomposition procedure enables voltage and current signals to be decomposed using two sets of different scaling and wavelet filters, of which one set has a narrow transition band and more stopband ripples, while the other set has wider transition bands and less ripples. 2) The proposed method is extended to both single-phase systems and three-phase systems, under either stationary or nonstationary, balanced or unbalanced situations. The rest of this paper is organized as follows. In Section \ref{sec2}, HT and its property are described, and the effect of interharmonics on the HT analysis is discussed. Section \ref{sec3} designs new scaling and wavelet filters, and proposes a two-stage decomposition method based on UWPT to extract each frequency component of signals. Instantaneous PQI definitions and the implementation procedure of estimating instantaneous PQIs are presented in Section \ref{sec4}. Two examples are taken in Section \ref{sec5} to test the performance of the proposed method. Section \ref{sec6} draws the conclusion.

\section{Estimation of Instantaneous Amplitudes and Phases via Hilbert Transform}
\label{sec2}
The first step of instantaneous PQIs detection is to estimate the instantaneous amplitudes and phases of voltage and current signals. The HT provides a linear convolution operation that introduces a phase shift of $-\pi/2$ at each positive frequency and $+\pi/2$ at each negative frequency. The HT of a real-valued signal $x(t)$ is defined as [23]
\begin{equation}
     \widetilde{x}(t)=\mathcal{H}[x(t)]=x(t)\ast\dfrac{1}{\pi t},
\end{equation}
where $\ast$ denotes the time convolution. The way to estimate instantaneous quantities of a signal is through its analytic signal defined as 
\begin{equation}
     \widehat{x}(t)=x(t)+j\cdot\widetilde{x}(t)=\left| \widehat{x}(t)\right|\cdot e^{j\varphi(t)},
\end{equation}
where $\left| \widehat{x}(t) \right|$ and $\varphi(t)$ are respectively the instantaneous amplitude and the instantaneous phase for the instant $t$. They are estimated by
\begin{equation}
    \left| \widehat{x}(t) \right|=\sqrt{x^2(t)+\widetilde{x}^2(t)},~\varphi(t)=\text{tan}^{-1}\left(\dfrac{\widetilde{x}(t)}{x(t)}\right).
    \label{eq3}
\end{equation}

\begin{figure}[!t]
\centering
\includegraphics[width=3.5in]{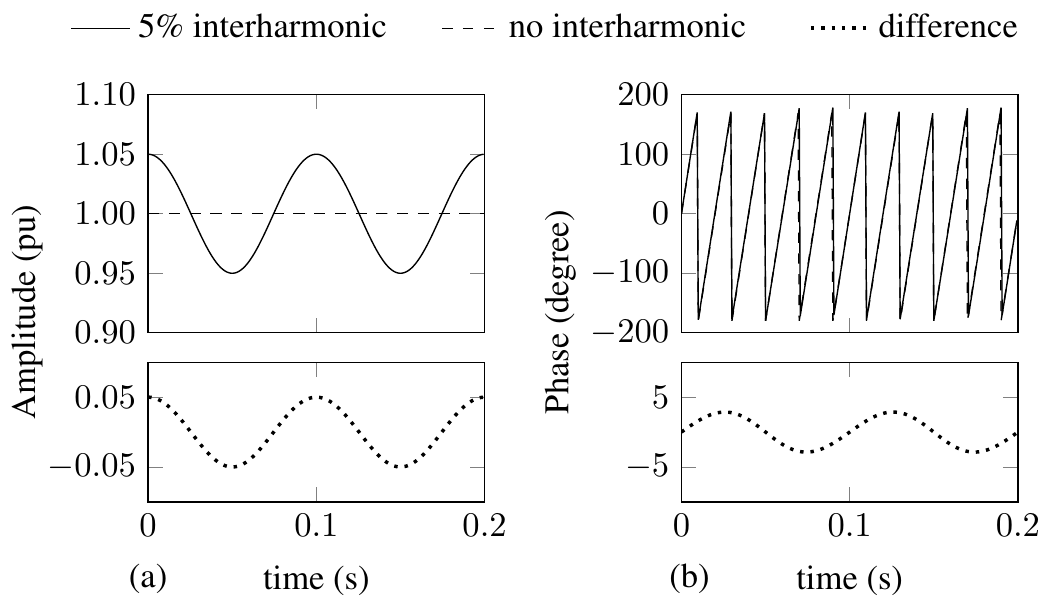}
\caption{The instantaneous amplitude and the instantaneous phase of a single-tone and a double-tone signal represented by HT. (a) shows the instantaneous amplitude (normalized to the fundamental component) and (b) presents the instantaneous phase variation over $0\leq t\leq 0.2$\,s.}
\label{fig1}
\end{figure}

It is worth pointing out that these instantaneous amplitudes and phases are accurate only when $x(t)$ is a single-tone signal. A major concern of this paper is the HT performance when $x(t)$ contains interharmonics. To test it, we demonstrated two HT examples: 1) $x(t)$ contains only the 50\,Hz fundamental component and 2) $x(t)$ contains 5\% additional interharmonic component at 60\,Hz. The signal of the latter case is written as
\begin{equation}
     x(t)=\sin(2\pi\cdot50t)+0.05\sin(2\pi\cdot60t).
\end{equation}
Using HT, the instantaneous amplitudes and the instantaneous phases of $x(t)$ during 0-0.2\,s are shown in Fig. \ref{fig1}. As it can be seen from the calculation results, the HT works perfectly for parameter estimations of a single-frequency signal. However, when interfering components are present, the instantaneous quantities (amplitude and phase) of an individual frequency component cannot be estimated accurately. This would lead to a large estimation error in the next step pf calculating instantaneous PQIs. To solve this issue, a suitable decomposition should be performed, which is particularly useful when the interharmonic component is close to fundamental component or integer harmonics.

\section{Decomposition Method in the Presence of Interharmonics}
\label{sec3}
\subsection{Undecimated Wavelet Packet Transform}
\label{3a}

\begin{figure}[!t]
\centering
\includegraphics[width=3.3in]{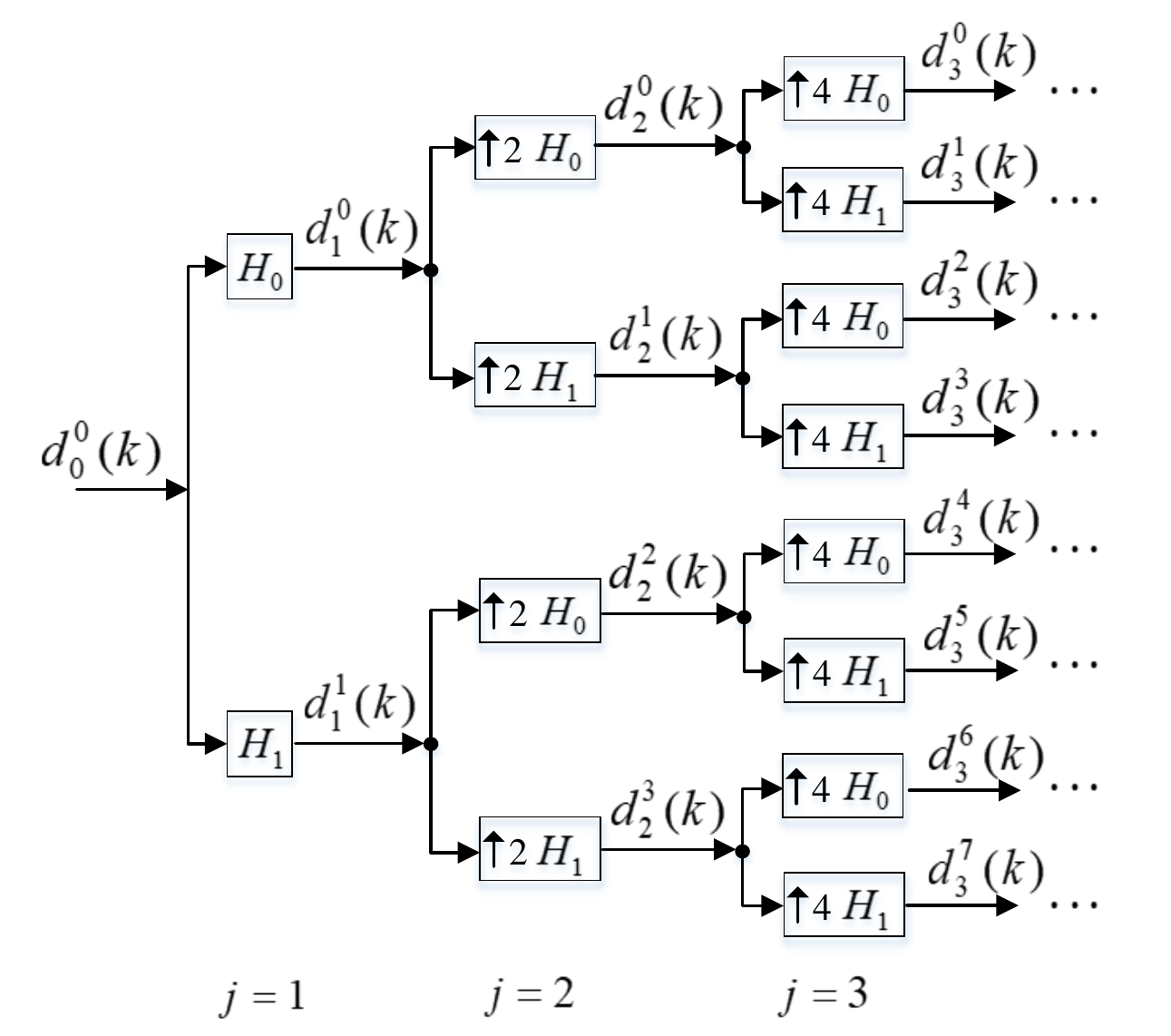}
\caption{Three level UWPT decomposition tree. $d_0^0(k)$ is the original sampled signal, $d_j^n(k)$ are wavelet packet coefficients of node $n$ at $j$th level. For each level of decomposition, previous level coefficients $d_{j-1}^n(k)$ are filtered into two equal-width frequency bands, { which belong} to $d_{j}^{2n}(k)$ and $d_{j}^{2n+1}(k)$, respectively.}
\label{fig2}
\end{figure}

Here we use UWPT as the tool to decompose different frequency components \cite{24}. In UWPT, the whole decomposition is realized using a set of low-pass and high-pass filters, $H_0(z)$ and $H_1(z)$, so-called respectively the scaling filter and the wavelet filter \cite{25}. In order to remain time { invariant}, UWPT carries out filter up-sampling instead of down-sampling by a factor of two on wavelet packet coefficients in WPT. As an example, the structure of a three-level UWPT is shown in Fig. \ref{fig2}.
At each decomposition level of UWPT, the scaling filter, $H_0(z)$, and the wavelet filter, $H_1(z)$, are up-sampled by padding zeros between each coefficient of the impulse responses of filters. Next, the decomposed wavelet packet coefficients are obtained by convolution of the previous level wavelet packet coefficients with up-sampled scaling and wavelet filters \cite{24}, as shown in Fig. \ref{fig2}. Wavelet packet coefficients at each level and each node reserve the same length as the original sampled signal. The wavelet packet coefficients at level $j$ and node $n$ in the decomposition tree can be calculated as follows:
\begin{eqnarray}
     d_{j}^{2n}(k)&=&d_{j-1}^n(k)\ast h_{0j}(k), \nonumber\\
     d_{j}^{2n+1}(k)&=&d_{j-1}^n(k)\ast h_{1j}(k),
\end{eqnarray}
where $h_{0j}(k)$ and $h_{1j}(k)$ are the impulse responses of up-sampled scaling and wavelet filters of level $j$.

\begin{figure}[!t]
\centering
\includegraphics[width=3.5in]{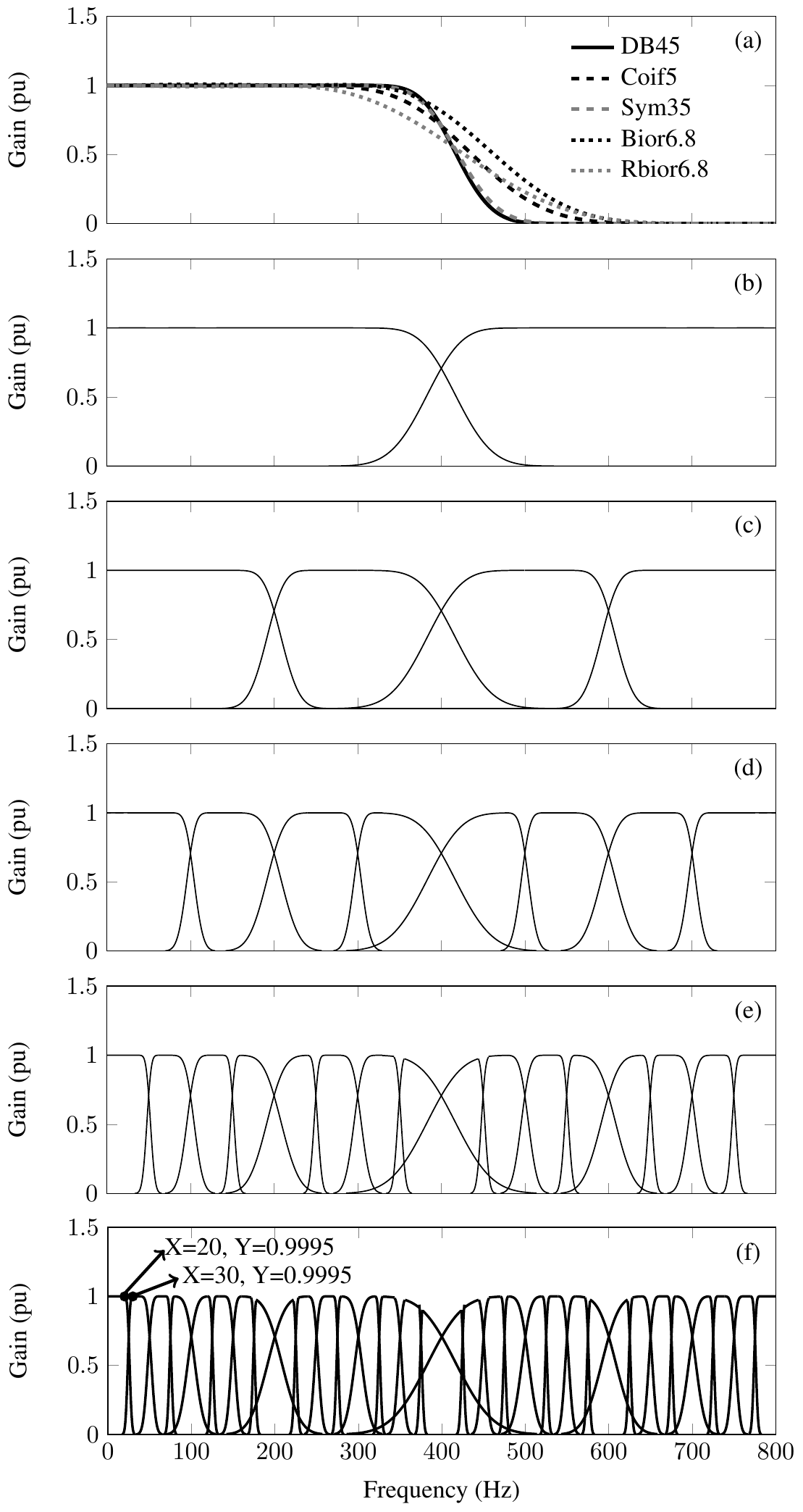}
\caption{Magnitude responses of scaling filters and wavelet filters based on conventional mother wavelets. The sampling frequency is 1600\,Hz and hence the cutoff is 800\,Hz. (a) Scaling filters at the first level of UWPT based on different mother wavelets. { (b)-(f) Frequency bands of UWPT from level one to level five based on ``DB45'' scaling filters and wavelet filters. The labels in (f) represent the transition bandwidth at the fifth level is 10\,Hz.}}
\label{fig3}
\end{figure}

However, similar to DWT and WPT, the major problem of UWPT is the spectrum leakage. When the signal { consists of} only the fundamental component and integer harmonics, the leakage problem does not exist in conventional wavelets (e.g., Daubechies wavelets) with frequency-shifting techniques \cite{17}. However, when the interharmonics are present, the conventional wavelets become inefficient to separate the adjacent frequency components. Fig. \ref{fig3}(a) shows the magnitude responses of scaling filters based on the mostly-used mother wavelets, i.e. ``DB45'', ``Coif5'', ``Sym35'', ``Bior6.8'', and ``Rbior6.8'' \cite{6,13,14,16,17}. As it can be seen, the ``DB45'' scaling filter has the narrowest transition band among the shown wavelets and its one- to five-level frequency bands of UWPT are shown in Fig. \ref{fig3}(b)-(f). The frequency bandwidth and the transition bandwidth are both halved as the decomposition level increases. It can be seen from Fig. \ref{fig3}(f) that even at the fifth decomposition level, the UWPT can only separate components with a frequency difference above 10\,Hz (transition bandwidth). { We also conduct a test on decomposing a multi-tone signal by different conventional wavelets. The test signal contains the 50\,Hz fundamental component and an additional 56\,Hz interharmonic component with 10\% of the fundamental amplitude. The maximum errors of 4.1\%, 23\%, 6.3\%, 37\%, and 23\% are given by ``DB45'', ``Coif5'', ``Sym35'', ``Bior6.8'', and ``Rbior6.8'' wavelets. These errors are significant, and it shows that conventional wavelets are not efficient for cases with interharmonics.}

Based on IEC 61000-4-7 \cite{8}, interharmonics may occur everywhere out of the harmonic subgroups (it has been assumed that no other frequency components are present between $f_h\pm5$\,Hz, other than the fundamental component and integer harmonics) \cite{26}. Therefore, the five-level UWPT based on conventional wavelets can separate interharmonics that are more than 10\,Hz away from the fundamental component or integer harmonics. For further resolution improvements, deeper decomposition is required but it needs a longer sampling data length and would cause heavier computation burdens. To address the above issues, an { optimized} solution is to { design} scaling and wavelet filters with narrower transition bands. We will present such a filter design in the following subsection.  

\subsection{New Scaling and Wavelet Filters Design}
\label{sec3.B}
According to the correspondence relationship between the filters of UWPT and the conjugate quadrature mirror filter bank (CQMFB), the process of designing CQMFB can be used to generate new scaling and wavelet filters \cite{27}.  

The low-pass filter $H_0(z)$ of CQMFB satisfies the following property:
\begin{equation}
     P(z)=H_0(z)H_0(z^{-1}),
\end{equation}
where $P(z)$ is a non-negative half-band filter, whose passband ripple equals to its stopband ripple. Since the passband cutoff frequency $\omega_\text{p}$ and stopband cutoff frequency $\omega_\text{s}$ of a half-band filter are equidistant from $\omega=\pi/2$, we can use the Chebyshev approximation and interpolation to firstly { obtain} an odd-order half-band filter $H_\text{LF}(z)$. Then $H_\text{LF}(z)$ is converted into the desired non-negative half-band filter $P(z)$ following
\begin{equation}
     P(z)=\dfrac{0.5}{0.5+\left|\delta\right|}H_\text{LF}^+(z),
\end{equation}
where $\left|\delta\right|$ denotes the maximum stopband ripple value of $H_\text{LF}(z)$ and $H_\text{LF}^+(z)=H_\text{LF}(z)+\left|\delta\right|$.

Next, $H_0(z)$ is obtained through the spectral decomposition of $P(z)$. $P(z)$ have $N-1$ ($N$ is the filter order) zeros which are conjugated in pairs and appear symmetrically with the unit circle as a mirror image. We can assign the zeros inside the unit circle to $H_0(z)$ and zeros outside the unit circle are assigned to $H_0(z^{-1})$. Obtaining $H_0(z)$, the high-pass filter $H_1(z)$ of CQMFB can be solved as
\begin{equation}
     H_1(z)=z^{-(N_\text{FB}-1)}H_0(-z^{-1}),
\end{equation}
where $N_\text{FB}$ is the order of $H_0(z)$ and $H_1(z)$.

\begin{figure}[!t]
\centering
\includegraphics[width=3.5in]{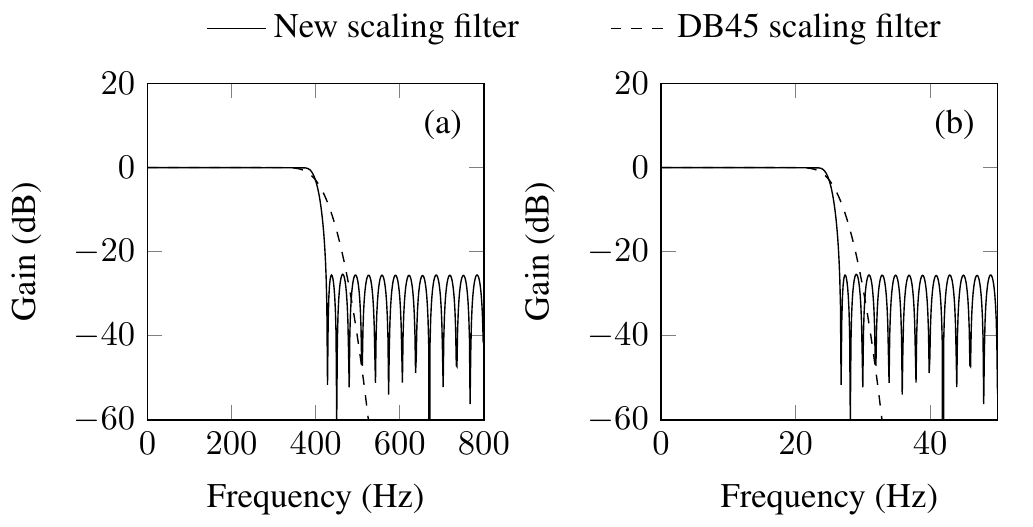}
\caption{Magnitude responses of the new scaling filter and ``DB45'' scaling filter. The sampling frequency is 1600\,Hz and the cutoff is 800\,Hz. (a) Scaling filters at the first level of UWPT. (b) Scaling filters at the fifth level of UWPT.}
\label{fig4}
\end{figure}

\begin{figure}[!t]
\centering
\includegraphics[width=3.5in]{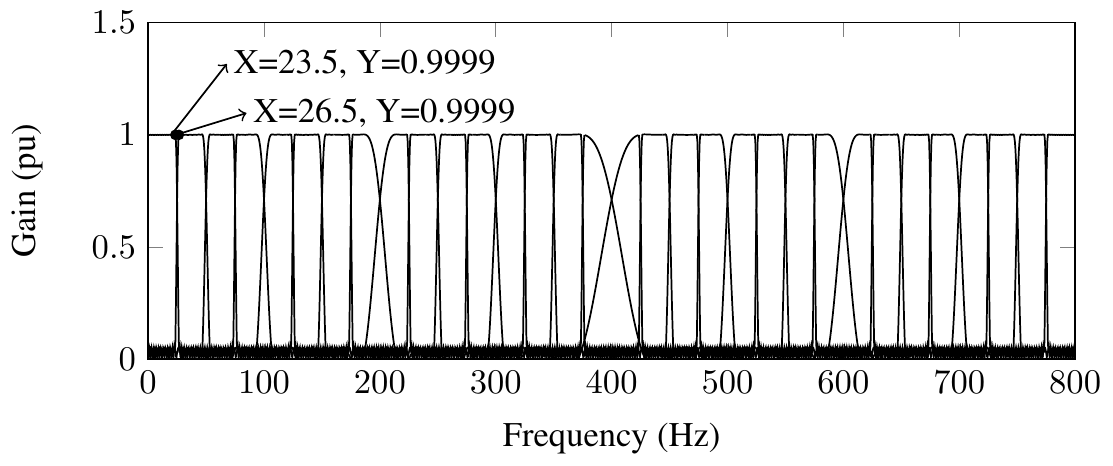}
\caption{Frequency bands of the five-level UWPT using the new scaling and wavelet filters. The sampling frequency is 1600\,Hz and the cutoff is 800\,Hz. The labels represent the transition bandwidth at the fifth level is 3\,Hz.}
\label{fig5}
\end{figure}

The low-pass and high-pass filters of CQMFB are used as the first-level scaling and wavelet filters in UWPT. Through a large number of simulations, we set the order of $P(z)$ to be $N=99$ so that the order of the scaling filter and the wavelet filter is $N_\text{FB}=50$, and set $\omega_\text{p}=0.47\pi$ to get a narrow filter transition bandwidth and fast stopband attenuation. In Fig. \ref{fig4}, the magnitude responses of the new scaling filter and ``DB45'' scaling filter at the first level and the fifth level of UWPT are shown. It can be seen that the new scaling filter has a narrower transition band than that of the ``DB45'' scaling filter, but with a higher stopband ripple. Fig. \ref{fig5} presents the frequency bands of the five-level UWPT decomposition using the new scaling and wavelet filters. In this case, two frequency components that are $\geq$3\,Hz away from each other can be well separated. Besides, The narrower transition band locates at multiple positions, i.e. [$(2l+1)\cdot25-1.5$, $(2l+1)\cdot25+1.5$] Hz, $l$ = 0, 1, 2, $\cdots$. This allows a more efficient separation of interharmonics that are close to the fundamental component and integer harmonics in both voltage and current signals. { For a simple test, here the example discussed in subsection \ref{3a}, i.e. 50\,Hz fundamental plus 10\% 56\,Hz interharmonic, is realized using the new scaling and wavelet filters. The maximum error is found only 0.2\%, which is more than 20\,times lower than the error of the best conventional wavelet, ``DB45''.} 

\subsection{Two-Stage Decomposition Method in the Presence of Interharmonics}
\label{sec3.C}
The above proposed scaling and wavelet filters can improve the frequency resolution for PQI detection { by narrowing the transition band}, but the stopband ripples may lead to an undesired error for decomposition of voltage and current signals. To compensate this error, a two-stage UWPT decomposition method is proposed as follows. In the first stage, the interharmonics of the voltage and current signals are estimated and removed from the original signals. This stage includes 6 steps, i.e.
\begin{enumerate}
    \item Estimate the frequency of interharmonics $f_i$ using the Hanning window-based two-point IpDFT \cite{28}.
    \item Find out the nearest harmonic frequency $f_h$ (or fundamental frequency) to each interharmonic frequency $f_i$. For example, the nearest harmonic frequency to 141 Hz interharmonic is $f_h=$ 150\,Hz, and the nearest harmonic frequency to 260 Hz interharmonic is $f_h=$\,250 Hz.
    \item Shift the frequency spectrum of original signal using single-sideband modulation (SSM) introduced in \cite{19}. { Note that the shifting frequency $f_\text{SSM}$ is an adaptive parameter to the analyzed signal,  because it is calculated according to the estimated interharmonic frequency:}
\begin{equation}
\left\{
             \begin{array}{lcl}
             \dfrac{f_h+f_i}{2}+f_\text{SSM}=f_h+25\,\mbox{Hz}~~f_h<f_i \\
             \dfrac{f_h+f_i}{2}+f_\text{SSM}=f_h-25\,\mbox{Hz}~~f_h>f_i 
             \end{array}  
\right.
\end{equation}
where $f_h$ = 50, 100, $\cdots$,\,Hz, and $\left|f_\text{SSM}\right|$ should be as small as possible. When $f_h<f_i$, $f_\text{SSM}>0$ and the frequency spectrum is shifted to the positive direction. Otherwise, $f_\text{SSM}<0$ and the frequency spectrum shift is toward the negative direction. 

\item Separate the interharmonic component from other frequency components by UWPT based on proposed scaling and wavelet filters in subsection \ref{sec3.B}.
\item Shift the interharmonic component back to its original frequency spectrum using SSM and subtract it from the original signal.
\item Repeat step 2) to 5) for each interharmonic.
\end{enumerate}

In the second stage, the voltage and current signals containing fundamental component and integer harmonics are decomposed into uniform frequency bands by using the FS-DWT method \cite{17}. This method is resulted from DWT and SSM, which by using conventional wavelets, can well surpress the spectrum leakage problem.

\section{Analysis of Instantaneous Power Quality Indices}
\label{sec4}
{ 
In this paper, instantaneous PQIs of both single-phase and three-phase systems are considered. They are the RMS values of voltage and current, total harmonic distortion, active power, reactive power, apparent power, power factor, and power frequency. Note that in three-phase systems, some of the quantities, e.g. voltage, current, power, harmonic distortion, etc, are equivalent values considering balance and unbalance situations. Detailed expressions are given in the Appendix. Different from the Fourier analysis of a certain time interval in \cite{9}, the PQI quantities defined in this paper are instantaneous. The magnitudes and phases of each voltage/current frequency component are calculated by applying (3) at each node of the UWPT.

}


\begin{figure}[!t]
\centering
\includegraphics[width=3.5in]{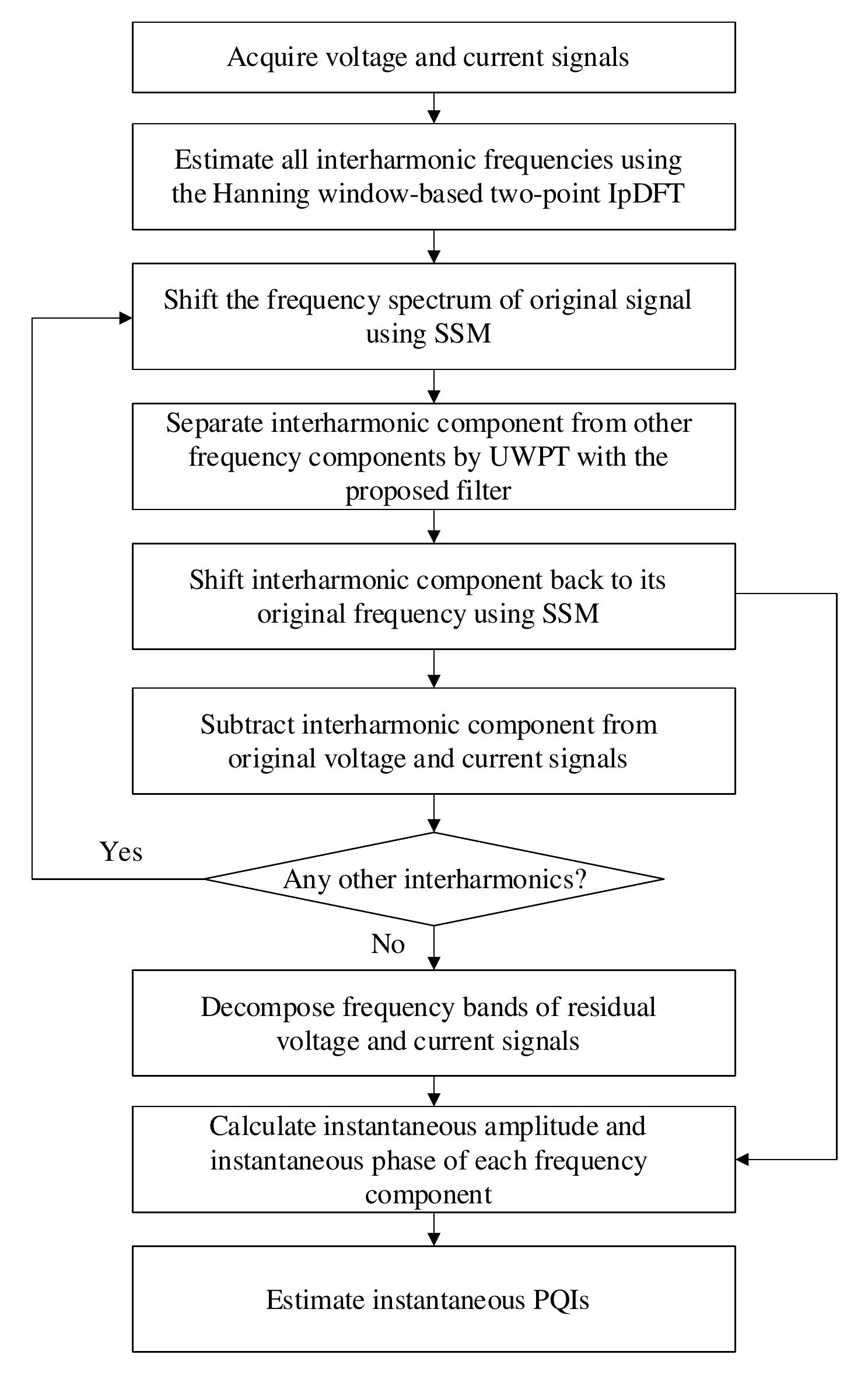}
\caption{Flowchart of estimating instantaneous PQIs.}
\label{fig6}
\end{figure}

The implementation procedure for estimating instantaneous PQIs is summarized and depicted in Fig. \ref{fig6}. In the first step, the voltage and current signals of a single-phase or three-phase system are acquired using a sampling frequency of $f_\text{s}$ ($f_\text{s}/50$ should be $2^M$, where $M$ is an integer) for 50-Hz system. Secondly, all the interharmonics contained in voltage and current signals should be removed using the first stage of the decomposition method described in subsection \ref{sec3.C}. This procedure needs to solve a $M$-level UWPT with the proposed filter. 
In order to reduce the computation burdens, filter coefficients of each node can be designed offline and stored in the memory in advance. Next, to obtain the fundamental component and harmonics, the FS-DWT method introduced by \cite{17} is applied to decompose the residual voltage and current signals, in which a conventional mother wavelet, e.g. ``DB40'', is used and the decomposition level is $M-1$. Similarly, filter coefficients of the lowest node are calculated offline and stored in the memory to fasten the calculation. After, the HT generates the analytic signal for each decomposed component. Note that the HT analysis also fits the interharmonic components that have been subtracted from original signals. The instantaneous amplitude and phase of each individual frequency component are estimated by (\ref{eq3}). Finally, the instantaneous PQIs are solved by using (\ref{eq13}) to (\ref{eq26}) { in the Appendix} and (10) to (15) in \cite{19}. 

{
The followings are the consideration of choosing the decomposition level in two stages: The IEC 61000-4-7 assumes that no other frequency components are present between $f_h\pm5$\,Hz other than the fundamental component and integer harmonics \cite{8}, and it therefore sets an maximum frequency resolution of 5\,Hz for interharmonic analysis. The decomposition level in the first UWPT is $M$, where the transition bandwidth of the proposed filter is already below 5\,Hz. Any further decomposition will not considerably improve the performance of the proposal. The second decomposition is to separate the fundamental and integer harmonics and the maximum frequency band should be lower than 50\,Hz. The $M-1$ or higher level decomposition will fulfill this requirement.
}

\section{Performance Tests}
\label{sec5}

Here we take two examples to test the performance of the proposed decomposition method for estimating instantaneous PQIs in systems with interharmonics and transient disturbances. In both cases, the length of the voltage and current signals are set to 4.6\,s. The sampling frequency is $f_\text{s}=6400\,\text{Hz}$, and the window length of 0.4\,s is chosen for each analysis. To minimize the HT end effect, an overlapped sliding-window process are conducted in the test.

As a comparison, another three methods, i.e. FS-DWT method \cite{17}, FS-WPT method \cite{19}, and STFT \cite{10}, are also demonstrated to calculate PQIs. In these tests, the STFT uses an observation window of 0.2\,s and can only produce one set of results per window. The FS-DWT method uses ``DB40'' as its mother wavelet, and the FS-WPT uses the mother wavelet presented in \cite{19}. { To make the test result comparable to FS-DWT, the same mother wavalet ``DB40'' is used in stage two of our proposed approach. Note that the mother wavelet is not fixed in this stage. Some other wavelets, e.g. ``DB45'', can offer better performance than ``DB40'', and hence can be employed for further enhancement of the proposed method. 
} 

\begin{table}[!t]
\centering
\renewcommand{\arraystretch}{1.3} 
\caption{Values of Magnitude, Frequency, and Initial Phase Angles for the Frequency Components of the Single-Phase Voltage and Current Signals} 
\begin{tabular}{|C{1.3cm}|C{1.3cm}|C{1.3cm}|C{1.3cm}|C{1.3cm}|}
\hline
\multirow{2}{*}{} & \multicolumn{2}{c|}{Voltage} & \multicolumn{2}{c|}{Current}  \\
\cline{2-5}
 Frequency (Hz) & Magnitude (V) & Phase angle ($\degree$) & Magnitude (A) & Phase angle ($\degree$) \\
\hline
 50.1 & 220 & 0 & 10 & -30 \\
\hline
 58 & 11 & 66 & 0.5 & 34 \\
\hline
 100.2 & 20 & 39 & 1.2 & 5 \\
\hline
 142 & 11 & 37 & 0.5 & 85 \\
\hline
 150.3 & 44 & 60.5 & 4 & 64 \\
\hline
 200.4 & 11 & 123 & 1 & 77 \\
\hline
 250.5 & 30 & -52 & 2.2 & 49 \\
\hline
 262 & 11 & 42 & 0.5 & 12 \\
\hline
 300.6 & 2 & 146 & 0.6 & 15 \\
\hline
 350.7 & 5 & 97 & 0.9 & 61 \\
\hline
 400.8 & 1 & 56 & 0.4 & 37 \\
\hline
 450.9 & 3 & 43 & 0.5 & 53 \\
\hline
\end{tabular}
\label{tab1}
\end{table}

\begin{table*}[!t]
\centering
\renewcommand{\arraystretch}{1.3} 
\caption{Transient Disturbances in the Single-Phase Voltage and Current Signals}
\begin{tabular}{|c|l|}
\hline
Starting time & \multicolumn{1}{c|}{Transient events} \\
\hline
$t_1=0.7$\,s & {Step change:} the system frequency $f_1$ changes {from 50.1\,Hz to 50\,Hz.} \\
\hline
$t_2=1.3$\,s & {{Step change:} the magnitudes of all current components increase 10$\%$.} \\
\hline
{$t_3=1.9$\,s} & {{Step change: the third harmonic disappears.}} \\
\hline
$t_4=2.09$\,s & {{A 900-Hz oscillating transient is superimposed onto current.}} \\
\hline
$t_5=2.1$\,s & {The oscillating transient disappears.} \\
\hline
{$t_6=2.7$\,s} & {{Step change: the third harmonic restores, and the second and fifth harmonics disappear.}} \\
\hline
$t_7=3.1$\,s & {{Step change:} the phase angle of current fundamental component changes from $-30\degree$ to $-20\degree$.} \\
\hline
$t_8=3.5$\,s & {{Step change: the magnitudes of all voltage and current components drop 10\%.}} \\
\hline
$t_9=4.1$\,s & {All harmonics and interharmonics disappear, {and the fundamental component remains stable.}} \\
\hline
\end{tabular}
\label{tab2}
\end{table*} 

\begin{figure*}
\center
\includegraphics[width=6.9in]{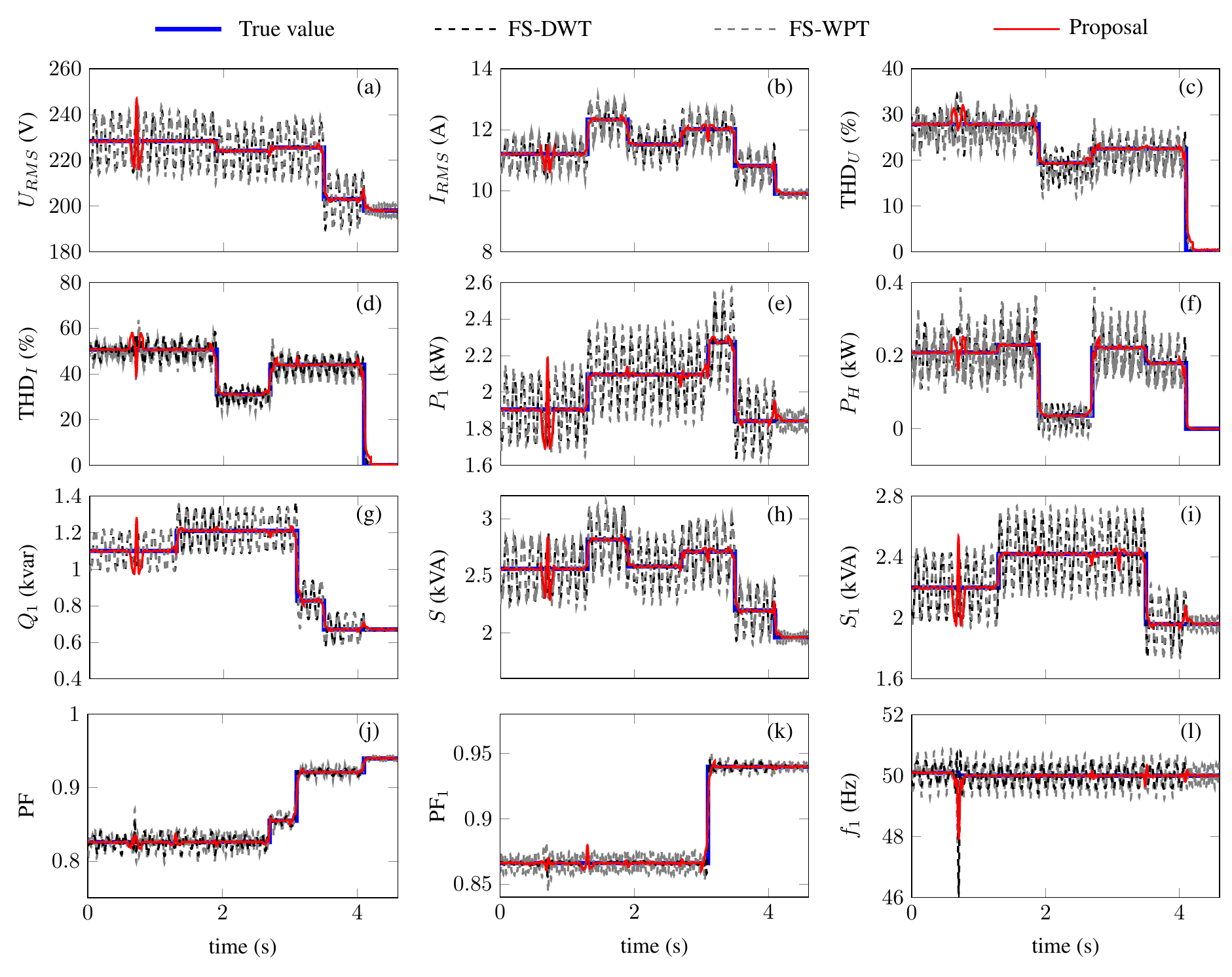}
\caption{Instantaneous PQIs estimation results for the single-phase system: (a) $U_\text{RMS}$, (b) $I_\text{RMS}$, (c) $\text{THD}_U$, (d) $\text{THD}_I$, (e) $P_1$, (f) $P_H$, (g) $Q_1$, (h) $S$, (i) $S_1$, (j) $\text{PF}$, (k) $\text{PF}_1$, and {(l) $f_1$}.}
\label{fig7}
\end{figure*}

\subsection{Single-Phase System}
The first example is a single-phase system. The voltage and current are multi-staged signals. Both signals contain twelve frequency components, including the fundamental component, 2nd-9th harmonics and three interharmonics (58\,Hz, 142\,Hz, and 262\,Hz). The magnitude, frequency, and initial phase angle of each frequency component for the starting stage are listed in Table \ref{tab1}. Nine transient disturbances, listed in Table \ref{tab2}, are considered during the test. { A white Gaussian noise with signal-to-noise ratio (SNR) of 40\,dB is injected into voltage and current signals.}

The estimated PQIs in the single-phase system are: voltage RMS value $U_\text{RMS}$, current RMS value $I_\text{RMS}$, total harmonic distortion of the voltage $\text{THD}_U$, total harmonic distortion of the current $\text{THD}_I$, fundamental active power $P_1$, harmonic active power $P_H$, fundamental reactive power $Q_1$, apparent power $S$, fundamental apparent power $S_1$, power factor $\text{PF}$, fundamental power factor $\text{PF}_1$, and { fundamental frequency $f_1$}. The instantaneous values of these PQIs calculated by the proposed, FS-DWT, and FS-WPT method are depicted in Fig. \ref{fig7}. Note that the true values are obtained based on the input signal. From Fig. \ref{fig7}, it can be seen that the proposed method yields a much more accurate estimation result than using FS-DWT and FS-WPT. For most stationary and transient conditions, the estimation accuracy of the FS-DWT and the FS-WPT method is considerably affected by interharmonics. The merit of the proposed method is that it can separate the interharmonics efficiently from the fundamental or integer harmonic components, and hence the estimation result is much less affected. On the overall trend, the proposed method can follow fast and accurately the real disturbances, but with a much lower estimation noise compared to the other two methods. It is worth to mention that the proposed method does not significantly improve the accuracy in detecting frequency transient compared to FS-DWT and FS-WPT. This can be seen by comparing the response results of the first transient event set in Table \ref{tab2}. The reason is that the frequency transient would lead to a spectrum aliasing and the leakage spectrum modulates with the nearby interharmonics, which will lead to a frequency resolution loss for the PQI detection. 

\begin{table}[!t] 
\centering
\renewcommand{\arraystretch}{1.3} 
\caption{Relative Errors of Estimation Results for the Single-Phase System [$\%$]} 
\begin{tabular}{|C{0.6cm}|C{0.6cm}|C{1.2cm}|C{1.2cm}|C{1.2cm}|C{1.2cm}|}
\hline
\textbf{PQIs} & \textbf{Stage} & \textbf{Proposal} & \textbf{STFT} & \textbf{FS-DWT} & \textbf{FS-WPT} \\
\hline
\multirow{3}{*}{$U_\text{RMS}$} & 1 & 0.0033 & 0.9709 & 0.8366 & 0.7607	\\
\cline{2-6} 
 & 2 & 0.0417 &	0.7899	&	0.7801	&	0.7056	\\
\cline{2-6} 
 & 3	&	0.0096	&	0.6575	&	1.1495	&	1.1616	\\
\hline
\multirow{3}{*}{$I_\text{RMS}$} & 1	&	0.0137	&	0.9861	&	0.9652	&	0.7487	\\
\cline{2-6} 
 & 2	&	0.0208	&	0.7020	&	0.5745	&	0.3707	\\
\cline{2-6} 
 & 3	&	0.0354	&	0.4244	&	0.7914	&	1.0298	\\
\hline
\multirow{3}{*}{$\text{THD}_U$} & 1	&	0.0519	&	5.1095	&	3.0349	&	4.6550	\\
\cline{2-6} 
 & 2	&	0.1939	&	6.3499	&	3.4711	&	7.0435	\\
\cline{2-6} 
 & 3	&	0.3301	&	2.9018	&	5.6661	&	7.3822	\\
\hline
\multirow{3}{*}{$\text{THD}_I$} & 1	&	0.0083	&	1.7080	&	0.4439	&	1.6376	\\
\cline{2-6} 
 & 2	&	0.1695	&	1.8023	&	3.8338	&	6.4462	\\
\cline{2-6} 
 & 3	&	0.0417	&	2.0995	&	0.0619	&	1.6201	\\
\hline
\multirow{3}{*}{$P_1$} & 1	&	0.0180	&	1.2336	&	2.1014	&	2.2093	\\
\cline{2-6} 
 & 2	&	0.0699	&	1.2222	&	1.8549	&	1.9348	\\
\cline{2-6} 
 & 3	&	0.0245	&	1.6374	&	1.6045	&	1.5079	\\
\hline
\multirow{3}{*}{$P_H$} & 1	&	0.0798	&	2.0343	&	0.3884	&	1.7551	\\
\cline{2-6} 
 & 2	&	0.3099	&	70.7991	&	21.4914	&	30.7266	\\
\cline{2-6} 
 & 3	&	0.2852	&	9.7622	&	5.1646	&	7.4422	\\
\hline
\multirow{3}{*}{$Q_1$} & 1	&	0.1434	&	1.3077	&	2.2902	&	2.2710	\\
\cline{2-6} 
 & 2	&	0.1278	&	0.7774	&	1.9756	&	1.9257	\\
\cline{2-6} 
 & 3	&	0.1218	&	1.0347	&	2.0032	&	1.9753	\\
\hline
\multirow{3}{*}{$S$} & 1	&	0.0171	&	1.9474	&	1.9372	&	1.6475	\\
\cline{2-6} 
 & 2	&	0.0625	&	1.4864	&	1.4625	&	1.1855	\\
\cline{2-6} 
 & 3	&	0.0450	&	1.0846	&	1.8200	&	2.0563	\\
\hline
\multirow{3}{*}{$S_1$} & 1	&	0.0225	&	1.2521	&	2.1488	&	2.2294	\\
\cline{2-6} 
 & 2	&	0.0844	&	1.1108	&	1.8852	&	1.9380	\\
\cline{2-6} 
 & 3	&	0.0359	&	1.5670	&	1.6481	&	1.5580	\\
\hline
\multirow{3}{*}{$\text{PF}$} & 1	&	0.0412	&	1.0566	&	0.0468	&	0.1183	\\
\cline{2-6} 
 & 2	&	0.0009	&	0.9257	&	0.0316	&	0.1532	\\
\cline{2-6} 
 & 3	&	0.0026	&	0.4531	&	0.1213	&	0.0085	\\
\hline
\multirow{3}{*}{$\text{PF}_1$} & 1	&	0.0405	&	0.0188	&	0.0409	&	0.0287	\\
\cline{2-6} 
 & 2	&	0.0145	&	0.1126	&	0.0263	&	0.0078	\\
\cline{2-6} 
 & 3	&	0.0114	&	0.0692	&	0.0631	&	0.0608	\\
\hline
\multirow{3}{*}{{$f_1$}} & {1}	&	{0.0040}	&	{0.0509}	&	{0.1560}	&	{0.1383}	\\
\cline{2-6} 
 & {2} &	{0.0041}	&	{0.0479}	&	{0.1320}	&	{0.1046}	\\
\cline{2-6} 
 & {3} &	{0.0034}	&	{0.4646}	&	{0.1501}	&	{0.1613}	\\
\hline
\end{tabular}
\label{tab3}
\end{table}

{ Note that as shown in Fig. 7, it is not necessary to cause changes for all quantities at a fixed transient event, because some PQIs are related to parameters of this transient and the result should show a step change, while the other PQIs are independent and in this case will remain stable, e.g., $\text{THD}_U$ remains the same value during the second transient event at $t_2=$1.3\,s.}

To quantify the accuracy of the proposal, the relative errors of estimation results under stationary conditions are given in Table \ref{tab3}. As STFT only gives one set of results per 0.2\,s, to be comparable, the instantaneous results of the proposal, FS-DWT, and FS-WPT are averaged with the same time length. Errors of three stationary stages, 0.3-0.5\,s, 2.3-2.5\,s, 3.6-3.8\,s, tagged respectively as stage 1, 2, 3, are compared in table \ref{tab3}. For each PQI, the relative errors are calculated by the following equation

\begin{equation}
     RE=\left| \dfrac{\text{PQI}_{\text{esti}}-\text{PQI}_{\text{true}}}{\text{PQI}_{\text{true}}}\right| \cdot 100\%,
\end{equation}
where $\text{PQI}_{\text{esti}}$ represents the result estimated by the proposed, STFT, FS-DWT, or FS-WPT method; $\text{PQI}_{\text{true}}$ represents the true value.

Table \ref{tab3} shows that the proposal gives the most accurate results among four methods. The maximum relative error of the proposed method does not exceed $0.5\%$, whereas the maximum errors of { $70.8\%$, $21.5\%$, and $30.7\%$ } are given by the STFT, FS-DWT, and FS-WPT methods, respectively. The PQI estimation accuracy is improved by more than one order of magnitude when the proposed method is employed.  For $P_1$, $S$, and $S_1$, the improvement is about two magnitudes.

\subsection{Three-Phase System}
The second example is a four-wire three-phase system ($f_1=50$\,Hz), of which the voltage and current data are given in \cite{30}. In addition to the setup in \cite{30}, interharmonics and transient disturbances, listed in Table \ref{tab4} and Table \ref{tab5}, are contained in the signals. We assign the 42\,Hz and 259\,Hz interharmonics with the positive sequence, while the 161\,Hz interharmonic negative \cite{31}. 
{ Similarly, both voltage and current signals contain a white Gaussian noise with SNR of 40\,dB.}

\begin{table}[!t]
\centering
\renewcommand{\arraystretch}{1.3} 
\caption{Percentage Values of Magnitude, Frequency, and Initial Phase Angles for the Interharmonics of the Three-Phase Voltage and Current Signals. Base Voltage: $U_{al}=272.45\ \text{V}$. Base Current: $I_{al}=133.92\ \text{A}$.} 
\begin{tabular}{|C{1.0cm}|C{0.5cm}|C{1.25cm}|C{1.1cm}|C{1.25cm}|C{1.1cm}|}
\hline
\multirow{2}{*}{} &  & \multicolumn{2}{c|}{Voltage} & \multicolumn{2}{c|}{Current}  \\
\cline{3-6}
 Frequency (Hz) & Phase & Percentage Magnitude ($\%$) & Phase angle ($\degree$) & Percentage Magnitude ($\%$) & Phase angle ($\degree$) \\
\hline
\multirow{3}{*}{42} & $a$ & 2.01 & 66 & 4.92 & 34  \\
\cline{2-6} 
 & $b$ & 2.03 & -50 & 4.95 & -66  \\
\cline{2-6}
 & $c$ & 1.99 & 178 & 4.07 & 161  \\
\hline
\multirow{3}{*}{161} & $a$ & 2.16 & 37.4 & 4.55 & 85  \\
\cline{2-6} 
 & $b$ & 2.14 & 156 & 4.23 & -157  \\
\cline{2-6}
 & $c$ & 2.17 & -85 & 4.28 & -40  \\
\hline
\multirow{3}{*}{259} & $a$ & 1.98 & 42 & 5 & 12  \\
\cline{2-6} 
 & $b$ & 2.01 & -88 & 4.87 & -98  \\
\cline{2-6}
 & $c$ & 1.96 & 170 & 5 & 143  \\
\hline
\end{tabular}
\label{tab4}
\end{table}

\begin{table}[!t]
\centering
\renewcommand{\arraystretch}{1.3} 
\caption{Transient Disturbances in the Three-Phase Voltage and Current Signals}
\begin{tabular}{|C{1.5cm}|C{6.5cm}|}
\hline
Starting time & Transient events \\
\hline
$t_1=1.5$\,s & \multicolumn{1}{m{6.5cm}|}{{Step change:} the magnitude of voltage of phase $c$ drops 10\%, and the magnitude of current of phase $c$ increases 10\%.} \\
\hline
$t_2=1.7$\,s & \multicolumn{1}{m{6.5cm}|}{{Step change:} the magnitudes of voltage and current of phase $c$ restore to original values.} \\
\hline
$t_3=2.7$\,s & \multicolumn{1}{m{6.5cm}|}{{Step change:} the magnitudes of three-phase currents drop 50\% simultaneously.} \\
\hline
$t_4=3.8$\,s & \multicolumn{1}{m{6.5cm}|}{{The fundamental component and 42\,Hz interharmonic remain (magnitudes step drop 10\% and phases lag 20\degree), and the other frequency components are filtered out.}} \\
\hline
\end{tabular}
\label{tab5}
\end{table}

\begin{figure*}
\center
\includegraphics[width=6.9in]{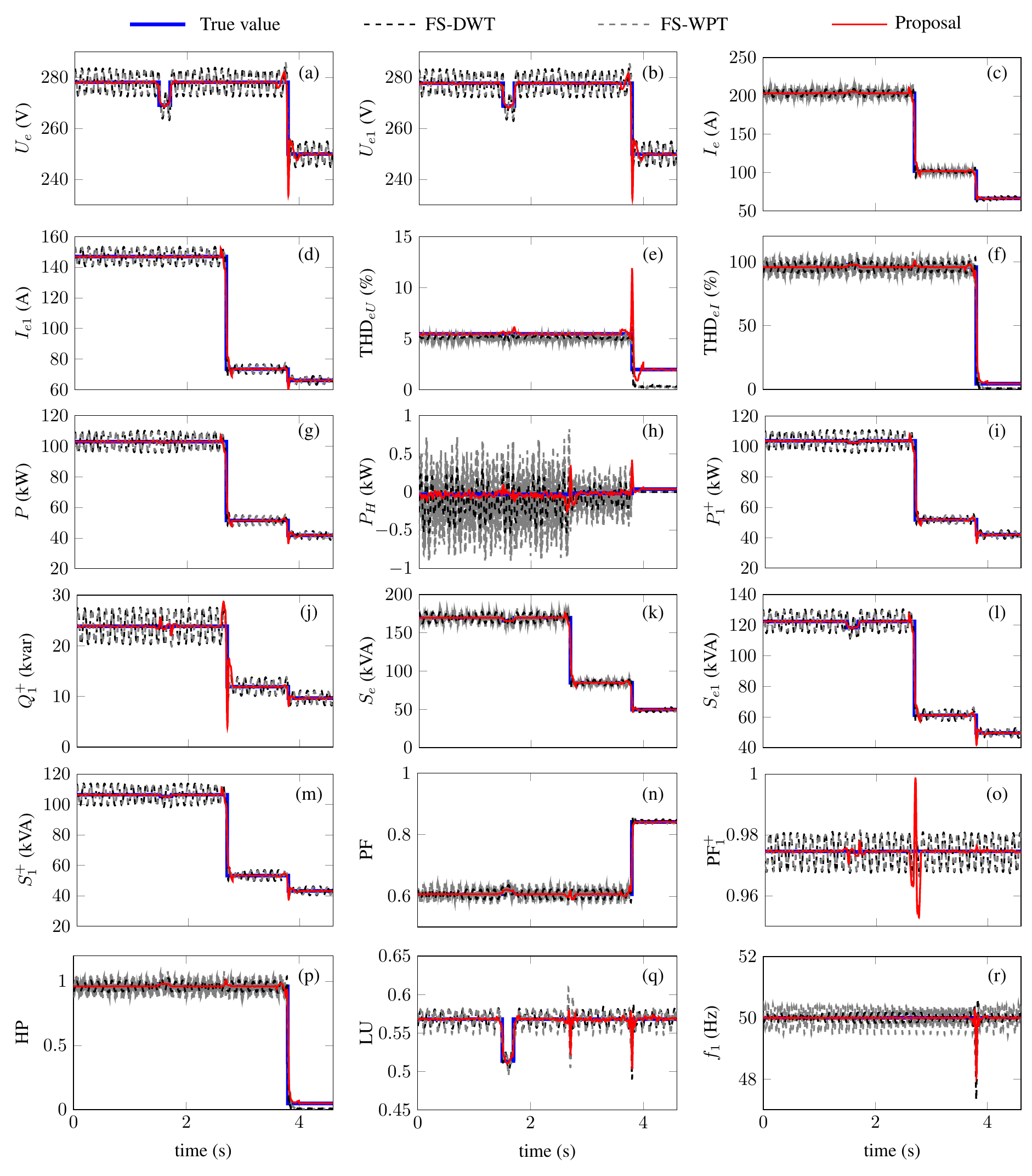}
\caption{Instantaneous PQIs estimation results for the three-phase system: (a) $U_e$, (b) $U_{e1}$, (c) $I_e$, (d) $I_{e1}$, (e) $\text{THD}_{eU}$, (f) $\text{THD}_{eI}$, (g) $P$, (h) $P_H$, (i) $P_1^+$, (j) $Q_1^+$, (k) $S_e$, (l) $S_{e1}$, (m) $S_1^+$, (n) $\text{PF}$, (o) $\text{PF}_1^+$, (p) $\text{HP}$, (q) $\text{LU}$, and {(r) $f_1$}.} 
\label{fig8}
\end{figure*}

Similar to the single-phase example, the instantaneous PQIs of the three-phase system are estimated by respectively the proposed, FS-DWT, FS-WPT methods, and the results are shown in Fig. \ref{fig8}. The estimated PQIs include $U_e$, $U_{e1}$, $I_e$, $I_{e1}$, $\text{THD}_{eU}$, $\text{THD}_{eI}$, $P$, $P_H$, $P_1^+$, $Q_1^+$, $S_e$, $S_{e1}$, $S_1^+$, $\text{PF}$, $\text{PF}_1^+$, $\text{HP}$, $\text{LU}$, and { the fundamental frequency $f_1$}, { which are given in the Appendix.} Note that during the calulation, we first extract the sequence component of each frequency contents, and then solve the PQIs related to these sequence components.

Again, the proposed PQI estimation approach is proven to be more efficient than the other two methods. 
It is interesting that at some PQI estimations, e.g., $Q_1^+$, $\text{PF}_1^+$, and $\text{LU}$, the proposed method leads to an adaptive response at $t=2.7$\,s where the magnitudes of three-phase currents drop 50$\%$ simultaneously. { It is found this reduced robustness} is caused by the unsynchronized reactions on three-phase currents when a step response is considered, which leads to a wrong estimation of the fundamental positive-sequence phase. 

{ Note that at $t_4=3.8$\,s, a large step change in voltage phases has a negative impact on $U_e$, $U_{e1}$, $\text{THD}_{eU}$, and $f_1$. Sparks of these PQI estimation results lead to a slow response to the transient event. However, in the reality, there will be a transition process when the active power filter (APF) is input, and the phases will not change dramatically and abruptly.}

\begin{table}[htbp] 
\centering
\renewcommand{\arraystretch}{1.3} 
\caption{Relative Errors of Estimation Results for the Three-Phase System [$\%$]} 
\begin{tabular}{|C{0.7cm}|C{0.6cm}|C{1.2cm}|C{1.2cm}|C{1.2cm}|C{1.2cm}|}
\hline
\textbf{PQIs} & \textbf{Stage} & \textbf{Proposal} & \textbf{STFT} & \textbf{FS-DWT} & \textbf{FS-WPT} \\
\hline
\multirow{3}{*}{$U_e$} &	1	&	0.0008	&	0.2043	&	0.3926	&	0.3509	\\
\cline{2-6} 
&	2	&	0.0207	&	0.2355	&	0.2385	&	0.2062	\\
\cline{2-6}
&	3	&	0.0207	&	0.2145	&	0.2597	&	0.2132	\\
\hline
\multirow{3}{*}{$U_{e1}$} &	1	&	0.0007	&	0.2493	&	0.3750	&	0.3279	\\
\cline{2-6}
&	2	&	0.0210	&	0.1916	&	0.2579	&	0.2312	\\
\cline{2-6}
&	3	&	0.0207	&	0.1723	&	0.2779	&	0.2373	\\
\hline
\multirow{3}{*}{$I_e$} &	1	&	0.0126	&	0.1880	&	0.4121	&	0.5663	\\
\cline{2-6}
&	2	&	0.0039	&	0.3739	&	0.2765	&	0.0406	\\
\cline{2-6}
&	3	&	0.0271	&	0.3172	&	0.3486	&	0.0570	\\
\hline
\multirow{3}{*}{$I_{e1}$} &	1	&	0.0226	&	0.1064	&	0.7530	&	0.6466	\\
\cline{2-6}
&	2	&	0.0305	&	0.3299	&	0.5661	&	0.4783	\\
\cline{2-6}
&	3	&	0.0154	&	0.2618	&	0.6685	&	0.4989	\\
\hline
\multirow{3}{*}{$\text{THD}_{eU}$} &	1	&	0.0633	&	16.3946	&	6.1032	&	8.0931	\\
\cline{2-6}
&	2	&	0.0901	&	16.0260	&	6.7167	&	8.8038	\\
\cline{2-6}
&	3	&	0.0022	&	15.3168	&	6.2934	&	8.4716	\\
\hline
\multirow{3}{*}{$\text{THD}_{eI}$} &	1	&	0.0211	&	0.6157	&	0.7771	&	0.1640	\\
\cline{2-6}
&	2	&	0.0556	&	0.0923	&	0.5387	&	0.9133	\\
\cline{2-6}
&	3	&	0.0245	&	0.1161	&	0.6044	&	0.9180	\\
\hline
\multirow{3}{*}{$P$} &	1	&	0.0265	&	2.0997	&	1.2246	&	1.0951	\\
\cline{2-6}
&	2	&	0.0330	&	1.5222	&	0.7573	&	0.6060	\\
\cline{2-6}
&	3	&	0.0661	&	1.5164	&	0.7834	&	0.5929	\\
\hline
\multirow{3}{*}{$P_H$} &	1	&	15.4524	&	2334.0	&	251.7	&	327.0	\\
\cline{2-6}
&	2	&	12.6341	&	3115.3	&	250.6	&	339.7	\\
\cline{2-6}
&	3	&	9.5467	&	3012.5	&	330.8	&	385.7	\\
\hline
\multirow{3}{*}{$P_1^+$} &	1	&	0.0256	&	2.7764	&	1.1456	&	1.0158	\\
\cline{2-6}
&	2	&	0.0340	&	0.5882	&	0.8126	&	0.6614	\\
\cline{2-6}
&	3	&	0.0634	&	0.6025	&	0.8616	&	0.6670	\\
\hline
\multirow{3}{*}{$Q_1^+$} &	1	&	0.1124	&	33.4024	&	1.8585	&	1.6743	\\
\cline{2-6}
&	2	&	0.1524	&	0.1228	&	2.8882	&	2.3936	\\
\cline{2-6}
&	3	&	0.0104	&	0.7849	&	3.0607	&	2.6681	\\
\hline
\multirow{3}{*}{$S_e$} &	1	&	0.0117	&	0.0160	&	0.7831	&	0.9018	\\
\cline{2-6}
&	2	&	0.0247	&	0.6085	&	0.5369	&	0.2629	\\
\cline{2-6}
&	3	&	0.0064	&	0.5310	&	0.6301	&	0.2861	\\
\hline
\multirow{3}{*}{$S_{e1}$} &	1	&	0.0220	&	0.3560	&	1.0883	&	0.9461	\\
\cline{2-6}
&	2	&	0.0516	&	0.5209	&	0.8657	&	0.7407	\\
\cline{2-6}
&	3	&	0.0053	&	0.4337	&	0.9872	&	0.7677	\\
\hline
\multirow{3}{*}{$S_1^+$} &	1	&	0.0187	&	1.2684	&	1.1596	&	1.0327	\\
\cline{2-6}
&	2	&	0.0399	&	0.5647	&	0.9368	&	0.7631	\\
\cline{2-6}
&	3	&	0.0598	&	0.5323	&	0.9928	&	0.7824	\\
\hline
\multirow{3}{*}{$\text{PF}$} &	1	&	0.0148	&	2.0834	&	0.4946	&	0.1878	\\
\cline{2-6}
&	2	&	0.0083	&	0.9193	&	0.1638	&	0.3427	\\
\cline{2-6}
&	3	&	0.0726	&	0.9907	&	0.0983	&	0.3053	\\
\hline
\multirow{3}{*}{$\text{PF}_1^+$} &	1	&	0.0069	&	1.4891	&	0.0165	&	0.0185	\\
\cline{2-6}
&	2	&	0.0060	&	0.0236	&	0.1217	&	0.0997	\\
\cline{2-6}
&	3	&	0.0037	&	0.0705	&	0.1286	&	0.1141	\\
\hline
\multirow{3}{*}{$\text{HP}$} &	1	&	0.0214	&	0.7073	&	0.7381	&	0.1159	\\
\cline{2-6}
&	2	&	0.0559	&	0.1837	&	0.5771	&	0.9620	\\
\cline{2-6}
&	3	&	0.0243	&	0.2037	&	0.6401	&	0.9647	\\
\hline
\multirow{3}{*}{$\text{LU}$} &	1	&	0.0135	&	3.7470	&	0.3513	&	0.3913	\\
\cline{2-6}
&	2	&	0.0476	&	0.1808	&	0.2286	&	0.0547	\\
\cline{2-6}
&	3	&	0.2234	&	0.4057	&	0.0379	&	0.0258	\\
\hline
\multirow{3}{*}{{$f_1$}} & {1}	& {0.0020}	& {0.1937}	& {0.0597}	& {0.0615}	\\
\cline{2-6}
&	{2}	& {0.0006}	& {0.0233}	& {0.0483}	& {0.0635}	\\
\cline{2-6}
& {3}	& {0.0013}	& {0.0130}	& {0.0503}	& {0.0613}	\\
\hline
\end{tabular}
\label{tab6}
\end{table}

The relative errors of estimation results under stationary conditions are given in Table \ref{tab6}. In this case, the durations 1, 2, 3 are 0.4-0.6\,s, 2.1-2.3\,s, and 3.1-3.3\,s. It is noticed that among all the PQI estimations, the harmonic active power $P_H$ exhibits the maximum error. The main reason is the spectrum leakage with interhamonics. Compared to STFT, relative errors of the FS-DWT and FS-WPT methods are less owing to the average process. However, these errors are still over 100\%. The proposed method can significantly reduce the $P_H$ error, to about 15\% level. For the other PQIs, the maximum relative error of the proposal is $0.15\%$ for $Q_1^+$, whereas the maximum errors of {$33.4\%$, $6.7\%$, and $8.8\%$} are given by the STFT, FS-DWT, and FS-WPT methods. For these parameters, the estimation accuracy is improved about one order of magnitude by the proposal. 

\subsection{Computational Time}

\begin{table}[!t]
\centering
\renewcommand{\arraystretch}{1.3} 
\caption{Computational Time of Three Methods}
\begin{tabular}{|C{1.5cm}|C{1.5cm}|C{1.5cm}|}
\hline
Case & Method & Time (s) \\
\hline
\multirow{3}{*}{Single-phase} & Proposal & 0.0536 \\
\cline{2-3}
  & FS-DWT & 0.0352 \\
 \cline{2-3}
  & FS-WPT & 0.4068 \\
\hline
\multirow{3}{*}{Three-phase} & Proposal & 0.3320 \\
\cline{2-3}
  & FS-DWT & 0.2874 \\
 \cline{2-3}
  & FS-WPT & 1.3240 \\
\hline
\end{tabular}
\label{tab7}
\end{table}

The calculation speed and the consumption of computer resources are also important measures of the PQI analysis algorithm. Here we evaluate the speed of the proposed method, compared to the FS-DWT and FS-WPT approaches. Table \ref{tab7} shows the computational time required to estimate instantaneous PQIs of one analysis window (0.4\,s). Each result is obtained by the average of 10 simulations. Due to the pre-processing of interharmonics, the proposal needs slightly more calculation time than the FS-DWT method. FS-WPT requires much more time than the other two methods. The main reason is that down-sampling are applied on wavelet packet coefficients in WPT, and the filter coefficients are not calculated and stored in advance. Although the proposal has a slower calculation speed than the FS-DWT, the speed decrease is not damaging (52\% and 16\% respectively for single-phase and three-phase systems), which enables the proposed method to be a suitable tool for instantaneous PQI estimation and monitoring.

\section{Conclusion}
\label{sec6}
In this paper, a decomposition method is proposed for PQIs assessment and monitoring in systems with interharmonics and transient disturbances. The proposed method is applicable for both single-phase systems and three-phase systems under general working conditions (stationary/nonstationary, balanced/unbalanced). The major improvement of the proposed estimation method, compared to other approaches, e.g., FS-DWT, FS-WPT, is the suppression of interharmonics. To achieve that a set of new scaling filter and wavelet filter with narrow transition bands are designed for UWPT. It is shown in both theoretical analysis and numerical simulations, these filters can efficiently separate the interharmonics from the fundamental and integer harmonics. Further, a two-stage decomposition method for multi-tone voltage and current signals is proposed, and the instantaneous PQIs are estimated using HT and PQI definitions in IEEE 1459 and IEC 61000-4-30. To verify the theory, numerical tests were demonstrated. The test results show that the proposed method can track the transitory change in voltage and current signals with a much higher accuracy and a better robustness than conventional methods, such as STFT, FS-DWT, and FS-WPT. In summary, with the interharmonics and transient disturbances more and more present in modern power systems, the proposal in this paper is expected to be a useful tool for PQI monitoring and PQ evaluations.

{  As is seen from Figs. \ref{fig7} and \ref{fig8}, a considerable noise appears at some transient events. How to remove these sparks should be investigated in the future. Some other performance improvements of the proposed method, such as fastening the response speed to different transient events, implementation under real complex conditions, are also considered.}

\appendices
\section*{Appendix}
\label{A}
\setcounter{equation}{0}
\renewcommand{\theequation}{A\arabic{equation}}
Whereas the single-phase instantaneous PQIs are already introduced in \cite{19}, here we provide a brief introduction to the instantaneous PQIs of the three-phase system.
The physical model for the voltage of a general four-wire three-phase system, with considerations of nonsinusoidal and unbalanced situations, is written as
\begin{align}
     u_a(t)=&\sqrt{2}U_{a1}\sin(\omega t+\alpha_{a1})+\sqrt{2}\sum_{h=2}^{H}{U_{ah}\sin(h\omega t+\alpha_{ah})} \nonumber\\
                 &+\sqrt{2}\sum_{i}{U_{ai}\sin(i\omega t+\alpha_{ai})}, \\
     u_b(t)=&\sqrt{2}U_{b1}\sin(\omega t+\alpha_{b1})+\sqrt{2}\sum_{h=2}^{H}{U_{bh}\sin(h\omega t+\alpha_{bh})} \nonumber\\
                 &+\sqrt{2}\sum_{i}{U_{bi}\sin(i\omega t+\alpha_{bi})}, \\
     u_c(t)=&\sqrt{2}U_{c1}\sin(\omega t+\alpha_{c1})+\sqrt{2}\sum_{h=2}^{H}{U_{ch}\sin(h\omega t+\alpha_{ch})} \nonumber\\
                 &+\sqrt{2}\sum_{i}{U_{ci}\sin(i\omega t+\alpha_{ci})}, 
\end{align} 
where $U_{a1}$, $U_{ah}$, $U_{ai}$, $U_{b1}$, $U_{bh}$, $U_{bi}$, $U_{c1}$, $U_{ch}$, and $U_{ci}$ are the RMS magnitudes of fundamental, the $h$th harmonic and the $i$th interharmonic components of $a$, $b$, and $c$ line-to-neutral voltages; $\alpha_{a1}$, $\alpha_{ah}$, $\alpha_{ai}$, $\alpha_{b1}$, $\alpha_{bh}$, $\alpha_{bi}$, $\alpha_{c1}$, $\alpha_{ch}$, and $\alpha_{ci}$ are their initial phase angles; $\omega=2\pi f_1$ is the fundamental angular frequency; $H$ is the maximum harmonic order. The line currents, $i_a$, $i_b$ and $i_c$, are defined by similar expressions to the voltage signal. 

Note that under unbalanced conditions, either the magnitude of three line currents is different, or their initial phase angles are shifted not precisely $120\degree$. This unbalance may also occur to voltage signals. IEEE 1459 \cite{9} defines the effective voltage and current quantities to treat three phases as a whole. Considering harmonic, interharmonic distortions and unbalance simultaneously, the effective voltage, $U_e$, and the effective current, $I_e$, are given by
\begin{eqnarray}
     U_e^2&=&\frac{1}{18}\sum\limits_{h=1}^{H}{[3(U_{ah}^2+U_{bh}^2+U_{ch}^2)+U_{abh}^2+U_{bch}^2+U_{cah}^2]}   \nonumber\\
       &&+\frac{1}{18}\sum\limits_{i}{[3(U_{ai}^2+U_{bi}^2+U_{ci}^2)+U_{abi}^2+U_{bci}^2+U_{cai}^2]}, \nonumber\\
     \label{eq13}
     &&\\
     I_e^2&=&\frac{1}{3}\sum\limits_{h=1}^{H}{[I_{ah}^2+I_{bh}^2+I_{ch}^2+I_{nh}^2]}\nonumber\\
       &&+\frac{1}{3}\sum\limits_{i}{[I_{ai}^2+I_{bi}^2+I_{ci}^2+I_{ni}^2]},
\end{eqnarray}
where $U_{abh}$, $U_{bch}$, $U_{cah}$, $U_{abi}$, $U_{bci}$, and $U_{cai}$ are the RMS magnitudes of the $h$th harmonic and the $i$th interharmonic components of line-to-line voltages; $I_{ah}$, $I_{bh}$, $I_{ch}$, $I_{nh}$, $I_{ai}$, $I_{bi}$, $I_{ci}$, and $I_{ni}$ are the RMS magnitudes of the $h$th harmonic and the $i$th interharmonic components of line currents and the neutral current. 

In this paper, the afore-mentioned voltage and current magnitudes are calculated by applying (\ref{eq3}) at each node of the UWPT. Hence, they are instantaneous quantities (effective voltage and effective current). In addition, the instantaneous equivalent total harmonic distortion for voltage, $\text{THD}_{eU}$, and current, $\text{THD}_{eI}$, are respectively defined as
\begin{equation}
     \text{THD}_{eU}=\frac{U_{eH}}{U_{e1}}\cdot100\%,~
     \text{THD}_{eI}=\frac{I_{eH}}{I_{e1}}\cdot100\%,
\end{equation}
where $U_{eH}$ and $I_{eH}$ are the nonfundamental effective voltage and current at a instant; $U_{e1}$ and $I_{e1}$ are the fundamental effective voltage and current at the same instant.

It is emphasized that the fundamental positive-sequence power quantities are more concerned in the three-phase nonsinusoidal and unbalanced systems. The instantaneous fundamental positive-sequence active power $P_1^+$, reactive power $Q_1^+$, and apparent power $S_1^+$ are defined as
\begin{eqnarray}
\begin{cases}
     P_1^+=&3U_1^+I_1^+\cos\theta_1^+   \\
     Q_1^+=&3U_1^+I_1^+\sin\theta_1^+   \\
     S_1^+=&3U_1^+I_1^+
\end{cases}
\end{eqnarray}
where $U_1^+$, $I_1^+$, and $\theta_1^+$ are calculated by the symmetrical components method \cite{29}; $U_1^+\angle\alpha_1^+$ and $I_1^+\angle\beta_1^+$ are the fundamental positive-sequence components of voltage and current, respectively, and $\theta_1^+=\alpha_1^+-\beta_1^+$. Again, all the magnitudes and phase angles are instantaneous values defined by (\ref{eq3}).

The instantaneous fundamental positive-sequence power factor is defined by
\begin{equation}
     \text{PF}_1^+=\frac{P_1^+}{S_1^+}=\cos\theta_1^+.
\end{equation}

The instantaneous active power, $P$, and the instantaneous harmonic active power, $P_H$, are solved as
\begin{eqnarray}
     P&=&\sum_{x=a,b,c}{[\sum_{h=1}^{H}{U_{xh}I_{xh}\cos\theta_{xh}}+\sum_{i}{U_{xi}I_{xi}\cos\theta_{xi}}]},\nonumber\\
     && \\
     P_H&=&P-\sum_{x=a,b,c}{U_{x1}I_{x1}\cos\theta_{x1}},
\end{eqnarray}
where $\theta=\alpha-\beta$, and the subscripts associated with different phases and different frequencies; $\beta$ indicates the current phase angle.

The instantaneous effective apparent power, $S_e$, and the instantaneous fundamental effective apparent power, $S_{e1}$, are given as the product of the corresponding instantaneous effective voltage and current, which are
\begin{eqnarray}
     S_e&=&3U_eI_e,   \\
     S_{e1}&=&3U_{e1}I_{e1}.
\end{eqnarray}

In order to quantify the harmonic distortions, the instantaneous harmonic pollution factor is defined as
\begin{equation}
     \text{HP}=\frac{S_{eN}}{S_{e1}},
\end{equation}
where $S_{eN}$ is the instantaneous nonfundamental effective apparent power. It is given by 
\begin{equation}
     S_{eN}=\sqrt{S_e^2-S_{e1}^2}.
\end{equation}

The instantaneous load unbalance factor is 
\begin{equation}
     \text{LU}=\frac{S_{U1}}{S_1^+},
\end{equation}
where $S_{U1}$ is the instantaneous fundamental unbalanced power, which is given by 
\begin{equation}
     S_{U1}=\sqrt{S_{e1}^2-(S_1^+)^2}.
\end{equation}


Lastly, the instantaneous power factor is defined by \cite{9}
\begin{equation}
     \text{PF}=\frac{P}{S_e}. 
     \label{eq26}
\end{equation}

\end{document}